\documentclass[a4paper, english]{amsart}
\usepackage[foot]{amsaddr}
\usepackage[utf8]{inputenc} 
\usepackage[T1]{fontenc}

\usepackage{amsmath} 
\usepackage{amssymb} 
\usepackage{amsthm}
\usepackage{mathtools}

\usepackage{graphicx} 
\usepackage[colorlinks,urlcolor=blue,citecolor=blue,linkcolor=blue]{hyperref} 
\usepackage{booktabs}
\usepackage{tabularx}

\usepackage{natbib}
\setcitestyle{authoryear,round,citesep={;},aysep={,},yysep=f{;}}

\usepackage{dsfont}  
\usepackage[capitalise]{cleveref}

\newtheorem{prop}{Proposition}

\crefname{equation}{}{}

\newcommand{\E}{\mathbb{E}}
\newcommand{\R}{\mathbb{R}}
\newcommand{\Z}{\mathbb{Z}}
\newcommand{\pr}{\mathbb{P}}
\newcommand{\var}{\mathrm{Var}}
\newcommand{\ind}{\mathds{1}}

\newcommand{\NsimIS}{N_\mathrm{IS}}
\newcommand{\dx}[1]{\mathrm{d#1}}
\DeclareMathOperator{\re}{Re}
\DeclareMathOperator{\vecop}{vec}
\DeclareMathOperator{\softmax}{softmax}
\newcommand{\multi}{\mathcal{M}}
\newcommand{\norm}{\mathcal{N}}
\newcommand{\bdot}{\boldsymbol{\cdot}}

\author[Voldoire]{Théo Voldoire}
\address[1]{Statistics Department, Harvard University, United States of America.}
\author[Chopin]{Nicolas Chopin}
\address[2]{ENSAE, Institut Polytechnique de Paris, France.}
\author[Rateau]{Guillaume Rateau}
\address[3]{Commissariat Général au Développement Durable, France.}
\author[Ryder]{Robin J. Ryder}
\address[4]{Mathematics Department, Imperial College London, United Kingdom.}

\title{Saddlepoint Monte Carlo and its application to exact Ecological Inference}

\begin{document}

\begin{abstract}
Assuming $X$ is a random vector and $A$ a non-invertible matrix, one
sometimes need to perform inference while only having access to samples of
$Y=AX$. The corresponding likelihood  is  typically intractable. One may
still be able to perform exact Bayesian inference using a pseudo-marginal
sampler, but this requires an unbiased estimator of the intractable
likelihood.  

We propose saddlepoint Monte Carlo, a method for obtaining an unbiased estimate
of the density of $Y$ with very low variance, for any model belonging to an
exponential family. Our method relies on importance sampling of the characteristic function, with insights
brought by the standard saddlepoint approximation scheme with exponential
tilting.  We show that saddlepoint Monte Carlo makes it possible to perform
exact inference  on particularly challenging problems and datasets. We focus on
the ecological inference  problem, where one observes only aggregates at a fine
level. We present in particular a study of the carryover of votes between the
two rounds of  various French elections, using the finest available data
(number of votes for each candidate in about 60,000 polling stations over most
of the French territory). 

We show that existing, popular approximate methods for ecological inference can
lead to substantial bias, which saddlepoint Monte Carlo is immune from. We
also present original results for the 2024 legislative elections on political
centre-to-left and left-to-centre conversion rates when the far-right is
present in the second round. Finally, we discuss other exciting applications
for saddlepoint Monte Carlo, such as dealing with aggregate data in privacy or
inverse problems.
\end{abstract}

\maketitle

\section{Introduction and setting}

\subsection{Motivation}

Certain applications require training models without access to the whole data,
but rather, a censored or transformed version of them. One such case is aggregation: for
privacy reasons, individuals are grouped in larger units and researchers may
only access marginals of the coerced data. For example, we may not access how
one individual votes at different rounds of elections, but only the counts of
voters grouped in a polling station.

A natural consequence of this constraint is the need for sampling or optimizing
from marginal distributions, which is a standard but difficult task in
computational statistics. A working paradigm to circumvent the difficulty is
that of ``data augmentation'' \citep{tanner1987calculation}, where
practitioners construct an MCMC sampler (or an EM-type optimizer) which
alternates between sampling hidden variables (respectively computing their
distribution) given parameters, and sampling (respectively optimizing)
parameters given hidden variables. However, this paradigm falls short in cases
where parameters and hidden variables are strongly correlated, motivating the need
for directly evaluating the marginal likelihood of parameters, without having
to introduce hidden variables or a completed likelihood.

This work is set in a Bayesian, ``pseudo-marginal'' approach
\citep{andrieu2009pseudo}: although the likelihood is not available
analytically, we have access to an unbiased estimator of it, allowing us to
construct Markov chains that \textit{exactly} target the parameter posterior.
The strength of pseudo-marginal approaches is
that, although they rely on  an (unbiased) approximation of the likelihood, the resulting algorithms are exact in the Monte Carlo
sense and do not make any approximation. 
Any level of accuracy can be attained with a corresponding computational budget, and the goal
of this work is to propose low-variance estimators so that this budget remains
very small.

\subsection{Formalization of the problem}\label{sub:formal}

The problem we wish to tackle may be formalized as follows. We consider  a random
vector $X$ of dimension $d_X$, and a $d_Y\times d_X $ non-invertible matrix $A$; 
typically $d_Y \ll d_X$. We posit some parametric model
for $X$, but we observe only $Y=AX$. We wish to compute the likelihood
(density) of $Y$ at observation $y$. 

A simple application of this framework is inference for aggregated data, e.g.
$A=(1, \ldots, 1)$. The main application (and running example) of this paper is
the study of two-round elections, with $I$ (resp.~$J$) candidates at the first
(resp.~second) round. There $X\sim \multi(n, p)$, with $n$ the number of
voters, and $X$ is the flattened version of the $I\times J$ table that contains
the number of individuals that voted for candidate $i$ at the first round and
candidate $j$ at the second round, for each possible pair $(i, j)$, $1\leq
i \leq I$, $1\leq j \leq J$. In this case, we observe only the $I+J$
\emph{margins} of that table (i.e. the number of votes for each candidate, at
each round, corresponding to the row and column sums of the $I\times J$ table); thus $A$ is the matrix (with entries equal to either zero or one)
which transforms $X$ into these $d_Y=I+J$ aggregates.

In practice, we may consider many such $X^k$ (one for each polling
station $k$), and these $X^k$ may have different distributions $\multi(n^k, 
p^k)$. This running example is a particular instance of the ecological
inference problem, which may be generalized in several ways (e.g., having three
rounds), while still pertaining to the framework considered here.

To keep notations simple, we assume in the main text that $X$ and $Y=AX$ takes values in
$\Z^{d_X}$ and $\Z^{d_Y}$ respectively. (See \cref{app:continuous_case} on how to adapt our
derivations to the more general cases where $X$ and $Y$ are continuous, or a
mix between discrete and continuous). 

The method we develop in this paper requires that the parametric model for
$X$ belongs to an exponential family, and  that the characteristic function of $X$ is
tractable. This is the case for our running example, where $X$ is multinomial.

\subsection{Related works in ecological inference}

In traditional ecological inference (EI) methods, computational inference problems have for long been an issue to
practitioners, putting constraints on possible substantive modeling goals. We
first argue that a majority of works have had to use approximate models instead
of exact ones because of computational reasons. A majority of works
present $X$ as drawn in two steps, and describe the first step (e.g., the racial
composition of a county) as a multinomial; but then, because of computational
difficulties, the second step (e.g., voting behavior) is presented as a distribution which
is close to a multinomial, but which is not a multinomial --- for example a Dirichlet distribution
\citep{king1999binomial} or a truncated Gaussian distribution \citep{lewis2004}. In these works, instead of modeling $X$ as a sequence of
multinomials, they rather model the second step by describing
\textit{frequencies}, based on the justification that the two should behave
very similarly. Similarly, \cite{imai2008bayesian} propose a mixture of
Gaussians, which has the advantage of being non-parametric, but still
approximates an empirical frequency using a Gaussian distribution. 
\cite{wakefield2001ecological} describes many solutions, but when it comes to
large ecological tables, proposes to approximate $X \sim \mathcal{M}(n, p)$ by
a Gaussian distribution with the correct mean and variance. On the other hand,
some works have used exact inference schemes, but these have difficulty scaling
up when the dimensionality increases, and exact applications have remained in
the realm of small ecological tables (e.g., \citealp{gnaldi2018ecological} 
successfully use a binomial distribution for the second step because they
work on a table with only two columns).

Overall, the justification for working with frequencies instead of counts and
using approximate distributions such as Gaussian or Dirichlet instead of
multinomial may be sensible, but the issue is that these approximations have
not been evaluated for larger ecological tables against an exact inference scheme,
because such a scheme was unavailable at a reasonable cost. We will show that, even if the number of individuals per polling station is large ($n^k > 1000$), approximating a
multinomial distribution by a Gaussian distribution changes the inference outcome, and
 the two posterior distributions may not even overlap. 
An additional advantage of our scheme is that it works not only for the desired multinomial distribution but also for any other distribution with a tractable characteristic function conditional on
parameters; this removes the dependency of the method on the functional form.

Other methods in EI have been proposed, but their approximate nature or limited
use cases are easier to pinpoint. An important baseline solution in EI has for
long been ecological regression \citep{goodman1953ecological}, but it does not allow for certain dependencies which are prevalent when
modeling $X$ with ``contextual effects''. The frequentist section in
\cite{rosen2001bayesian} proposes a method of moments estimator, which provides
an estimator that falls outside the traditional method of maximum likelihood
estimation and incurs error in the finite population regime.
Similarly, developments on maximum entropy estimation have been justified with asymptotic results \citep{elff2008ignoramus, bernardini2020entropy}, but this procedure displays some error in the finite population
setting. 

We introduce our new methodology, which we dub saddlepoint Monte Carlo, in a
general setting in \cref{s:method}. We showcase its application to ecological
inference in \cref{s:application}, where we also present several real-data
applications on French elections.

\section{Saddlepoint Monte Carlo}\label{s:method}

\subsection{Importance sampling through characteristic functions}

\subsubsection{Principle}\label{ss:principle_is_cfs}

Let $f_X$ and $\varphi_X$ denote respectively the probability mass function and
the characteristic function of random variable $X$:
\begin{equation*}
  \varphi_X(z) \coloneqq \E\left[ \exp(i z^\top X) \right], 
  \quad \forall z\in \R^{d_X}.
\end{equation*}

The characteristic function of $Y=AX$ is easily obtained from
$\varphi_X$: $\varphi_{AX}(z) = \varphi_X(A^\top z)$ for $z\in\R^{d_Y}$. The 
inversion formula gives: 
\begin{align}
  f_{AX}(y) 
  & = \frac{1}{(2\pi)^{d_Y}} \int_{[-\pi, \pi]^{d_Y}} 
  \exp(-iz^\top y) \varphi_{AX}(z)\dx{z} \label{eq:density_AX}\\
  & = \frac{1}{(2\pi)^{d_Y}} \int_{[-\pi, \pi]^{d_Y}} 
  \re\left\{\exp(-i z^\top y) \varphi_{X}(A^\top z) \right\} \dx{z} \notag
\end{align}
where $\re\{\bdot\}$ stands for the real part of its argument.

We may thus approximate without bias this integral using importance sampling,
with samples $(Z_n)$ drawn from an instrumental distribution $q$:
\begin{equation}\label{eq:estimate_density}
  \hat{f}_{AX}(y) = \frac{1}{\NsimIS} \sum_{n=1}^{\NsimIS} 
  \frac{\eta(Z_n)}{q(Z_n)},
  \qquad Z_1,\dots, Z_{\NsimIS} \sim q
\end{equation}
and $\eta(z) \coloneqq \re\left\{ \exp(-i z^\top x) \varphi_X(A^\top
z)\right\} / (2\pi)^{d_Y}$.

In this paper, we consider two proposal distributions: first, the uniform
distribution over $[-\pi, \pi]^{d_Y}$, $q(z) =\ind_{[-\pi, \pi]^{d_Y}}(z) /
(2\pi)^{d_Y}$, mainly due to its simplicity; 
and, second,  a proposal distribution based
on a Gaussian approximation, which we develop in the next section.

\subsubsection{Gaussian approximation proposal}\label{sub:gaussian_proposal}

Assume that we are able to compute the expectation $\mu_X \coloneqq \E[X]$ and
variance $\Sigma_X \coloneqq \var(X)$ of $X$. Assume furthermore that $Y=AX$ has
a near Gaussian distribution, with expectation $\mu_Y=A\mu_X$, variance
$\Sigma_Y=A \Sigma_X A^\top$.  This will happen for instance in our ecological
inference running example, $X\sim \multi(n, p)$ whenever $n$ is large. Note
that, in that case, the distribution of some of the components of $X$ may be
far from Gaussian (e.g., when $p_j$ is close to 0 or 1), while all the components
$Y=AX$ may nonetheless still be near Gaussian, due to an aggregation effect.

In such a situation, we expect that $\varphi_{AX}(y) \approx
\varphi_{Y^\prime}(y)$,
with $Y^\prime\sim \norm(\mu_Y, \Sigma_Y)$. This quantity has a closed-form
expression:
\begin{equation*}
  \varphi_{Y^\prime}(y) = \exp\left\{ i y^\top \mu_Y - \frac 1 2 y^\top \Sigma_Y y
  \right\}.
\end{equation*}

Thus, one may rewrite \cref{eq:density_AX} as:
\begin{equation}\label{eq:is_identity_for_AX}
  f_{AX}(y) 
  = \frac{1}{(2\pi)^{d_Y}} \int_{-[\pi, \pi]^{d_Y}}
  \exp\left\{ i z^\top (\mu_Y-y) \right\} \\
  \frac{\varphi_{AX}(z)}{\varphi_{Y^\prime}(z)}
  \exp\left( - \frac 1 2 z^\top \Sigma_Y z \right) \dx{z} 
\end{equation}
which suggests  importance sampling would be efficient with the instrumental distribution $\norm(0_{d_Y},
\Sigma_Y^{-1})$. The corresponding estimator has expression
\cref{eq:estimate_density}, with $q$ the probability density of
$\norm(0_{d_Y}, \Sigma_Y^{-1})$.

\subsubsection{Further considerations}

The importance sampling approach developed above requires closed-form
expressions for $\varphi_X$, the characteristic function of $X$.  The Gaussian
proposal requires furthermore that the expectation and variance of $X$ are also
in closed-form. 

We may reduce the variance of the importance sampling estimates proposed above
by using RQMC (randomised quasi-Monte Carlo); see \cref{app:rqmc} for more
details and, e.g., Chap.~17 of \cite{practicalqmc} for an overview or RQMC.

We expect the Gaussian proposal to outperform the uniform proposal whenever the
distribution of $AX$ is nearly Gaussian; this point is assessed in
\cref{ss:summary} and \cref{sec:num}. In the next section, we present a way to further reduce the variance of the estimator. 

\subsection{Exponential tilting for marginal distributions}\label{sub:tilting}

\subsubsection{Basics of exponential tilting}

Exponential tilting is a parametric change of measure that may be defined (in our
context) as, for any $\rho \in \R^{d_X}$:
\begin{equation*}
  f_{X_\rho}(x) 
  \coloneqq \frac{\exp\left(\rho^\top x\right)}{{M_X (\rho)}}f_X(x)
\end{equation*}
where $M_X(\rho) \coloneqq  \E[\exp(\rho^\top X)]$ is  the moment generating
function of $X$.

Exponential tilting is particularly natural when the distribution of $X$ belongs
to an exponential family. In that case, all tilted distributions belong to the
same family. For instance, for $X\sim \multi(n, p)$, $X_\rho \sim \multi(n,
p_\rho)$, with $p_{\rho,j} \propto p_j e^{\rho_j}$. For more background on
exponential tilting, see ~\cite{Butler_book}.

In our context, we note that a tilting on $X$ may define a tilting on $AX$. That is:
\begin{align*}
	f_{AX_\rho}(y)
	 & = \sum_x f_{X_\rho}(x) \ind\{Ax=y\}                                \\
	 & = \sum_x \frac{e^{\rho^\top x}}{{M_X (\rho)}}f_X(x)  \ind\{Ax=y\}
\end{align*}
and, provided we take $\rho = A^\top\nu$ for some $\nu\in\R^{d_Y}$, then 
\begin{align*}
	f_{AX_\rho}(y)
	 & = \frac{e^{\nu^\top  y}}{{M_X (A^\top \nu)}} \sum_x f_X(x)  \ind\{Ax=y\} \\
	 & = \frac{e^{\nu^\top  y}}{{M_{AX} (\nu)}} f_{AX}(y).
\end{align*}

The identity opens up the possibility to estimate $f_{AX}(y)$ under a different
distribution for $X$ (namely the distribution of $X_\rho$). Provided $X$
belongs to an exponential family, then this different distribution will also
belong to that family, and calculations under it may proceed exactly along the
same lines. We exploit this extra degree of liberty in the next section.

\subsubsection{Exponential tilting and importance sampling}

We are now able to generalize our importance sampling
estimator to 
\begin{equation}\label{eq:generalised_estimate}
  \hat{f}_{AX}^\nu(y) \coloneqq 
  \frac{M_X(A^\top \nu)}{\exp\left(\nu^\top  y\right)} \hat
  f_{AX_\rho}(y),\quad \rho = A^\top \nu,
\end{equation}
where $\hat f_{AX_\rho}(y)$ is \cref{eq:estimate_density} but with the distribution
of $X$ replaced by that of $X_\rho$. (In particular, $\varphi_X$ becomes
$\varphi_{X_\rho}$, in the definition of $\eta$, and the expectation and
variance of $X$ are replaced by those of $X_\rho$ in the Gaussian proposal).

We want to choose $\nu\in\R^{d_Y}$ in a way that makes the
variance of~\cref{eq:generalised_estimate} as small as possible. We found the
following heuristic to work very well to that end. Set $\nu$ so that 
\begin{equation}\label{eq:saddlepoint}
  A \nabla \kappa_{X}(A^\top \nu) = y,
\end{equation}
where $\kappa_X:\rho\mapsto\log
M_X(\rho) $ is the cumulant function of $X$.

A simple way to justify this heuristic is to observe that it is
equivalent to choosing $\nu$ such that $AX_\rho$ (with $\rho=A^\top \nu$) has
expectation $y$. (Note that $\nabla \kappa_X(\rho) = \E[X_\rho]$.)
In this way, we cancel the first factor in the integrand 
of~\cref{eq:is_identity_for_AX}. The importance weight function is then 
$\varphi_{AX}/\varphi_{Y'}\approx 1$, which should lead to an importance
sampling estimator with very low variance. In other words, for a fixed $y$, we
tilt the distribution of $X$ so that $y$ is no longer in the tail of the
distribution of $AX$, but instead in its `centre'.
A more rigorous way to justify this heuristic is to relate our method to
saddle-point approximations, as we do in the next section.

In practice,~\cref{eq:saddlepoint} does not admit a closed-form
solution, but it may be solved numerically through Newton's algorithm. We observe that only 2 to 3 Newton iterations are required to
converge to the solution in most cases.

\subsection{Relationship to the saddlepoint approximation method}\label{ss:why_saddle}

Our approach is very close in spirit to the classical saddlepoint
approximation method \citep{Butler_book}, which we describe now.

It is standard to introduce that method as a technique to approximate the density of
the sum of $n$ univariate variables, $Y=\sum_{i=1}^n X_i$. We can recover this
particular case by taking $A=(1,\dots, 1)$, $d_X=n$, $d_Y=1$. The  saddlepoint
approximation is then derived through essentially the same steps as above: (i)
express the density of $Y=AX$ (or its probability mass function if $Y$ is
discrete) via the inversion formula, as in~\cref{eq:density_AX};  (ii) apply
exponential tilting, using formula~\cref{eq:saddlepoint} to choose the tilting
parameter;  (iii) replace $\varphi_{AX}$ with the characteristic function of a
Gaussian distribution which approximates the distribution of $AX$ in an
asymptotic manner.

The main difference between our method (which we call from now on `saddlepoint
Monte Carlo') and the classical saddlepoint approximation, is that the
latter is deterministic, whereas the former relies on importance sampling to
provide a stochastic, unbiased estimator. Note also that the connection between
the two methods is stronger when we use a Gaussian distribution as a proposal,
as described in \cref{sub:gaussian_proposal}, but there are cases where other
proposals (such as the uniform distribution) lead to better results, as we
shall see in our numerical study.

Our presentation describes four methods (Gaussian or Uniform proposal, with or
without tilting), but the default strategy we suggest is to use the Gaussian
proposal with tilting. This is supported by theoretical considerations in the
next section, and by numerical studies in \cref{ss:summary}.

\subsection{A supporting result}\label{sub:supporting_theory}

In this section, we prove that the (relative) variance of saddlepoint Monte
Carlo goes to zero as $n$ (the \emph{data} sample size) goes to infinity,
provided we use a Gaussian proposal and tilting. Our result is specific to
the multinomial model $X\sim \multi(n, p)$.

\begin{prop}\label{prop:cv}
	
	Let $X\sim\multi(n,p)$, $t\in\R^{d_{Y}}$, and assume $A$ and $p$
	are such that the application $\lambda:\R^{d_{Y}}\to\R^{d_{Y}}$,
	\begin{equation}
		\lambda:\,\eta\mapsto pA\xi(A^{T}\eta),\quad\xi(\rho):=pe^{\rho}/\|pe^{\rho}\|\label{eq:technical_cond}
	\end{equation}
	defines a bijection locally around $t$; that is, there exists a neighborhood
	$U$ of $t$, such that $\lambda$ restricted to $U$ defines a bijection
	between $U$ and $\lambda(U)$. 
	
	Then one has, for $y_{n}=\lfloor nt\rfloor$,
	\[
	\var\left[\frac{\hat{f}_{AX_{n}}^{\star}(y_{n})}{f_{AX_{n}}(y_{n})}\right]\to0, 
    \qquad\text{as }n\to\infty
	\]
	where $\hat{f}_{AX}^{\star}(y)$ denotes the tilted/Gaussian estimator,
    i.e., $\hat{f}_{AX}^{\star}(y) \coloneq \hat{f}_{AX}^{\eta^\star}(y)$ and
    $\eta^\star$ is the solution of \cref{eq:saddlepoint}.
\end{prop}

For a proof, see \cref{app:proof_convergence}. A few remarks are in order. 

First, this result holds for a \emph{fixed} number of simulations $\NsimIS$,
even for $\NsimIS=1$. What is says is that taking $n\to \infty$ makes the
distribution of $X$ and $Y=AX$ more and more Gaussian, and that this is
sufficient to achieve a zero relative error for the tilted saddlepoint Monte
Carlo estimator (based on a Gaussian proposal).

Second, trying to establish this result for a \emph{fixed} value $y$ does not
seem to work and would not make much sense: the probability $f_{AX_{n}}(y)$ is
exponentially small as $n\to\infty$, as this is the  probability of observing
fixed counts whereas the number of individuals $n$ goes in infinity. We take
instead $y_{n}=\lfloor nt\rfloor,$ so that $y_{n}/n\approx t$, but we need to
take the integer part since $AX$ takes values in $\mathbb{N}^{d_{Y}}$. This is
a technicality that is specific to the discrete case.

Third,  the technical condition~\cref{eq:technical_cond} is needed to avoid
pathological cases; for instance, if $A=(1,\dots,1)$, then $\lambda(\eta)=1$
for all $\eta$, so the function is not a bijection. In that case, we have
$Y=AX=n$ which is an uninteresting case.  The technical condition also allows
us to avoid having any component of $t$ equal to zero; note that if the
observed data do include some marginals equal to 0, then all the corresponding
entries in $X$ are necessarily $0$ also and we can reformulate the problem in a
lower dimension, in which condition \cref{eq:technical_cond} might be verified; this is the strategy we recommend in practice.

\subsection{Summary, numerical evaluation}\label{ss:summary}

Given an observation $y$ and a distribution for $X$ (which belongs to
an exponential family), we now have four ways to estimate  
$f_{AX}(y)$ without bias: 

\begin{itemize}
  \item We may use either a uniform proposal (\cref{ss:principle_is_cfs}) or a
    Gaussian proposal (\cref{sub:gaussian_proposal}) in our importance sampling
    scheme; 
  \item We may or may not use exponential tilting to further reduce the
    variance (\cref{sub:tilting}).
\end{itemize}

\begin{figure}
	\includegraphics{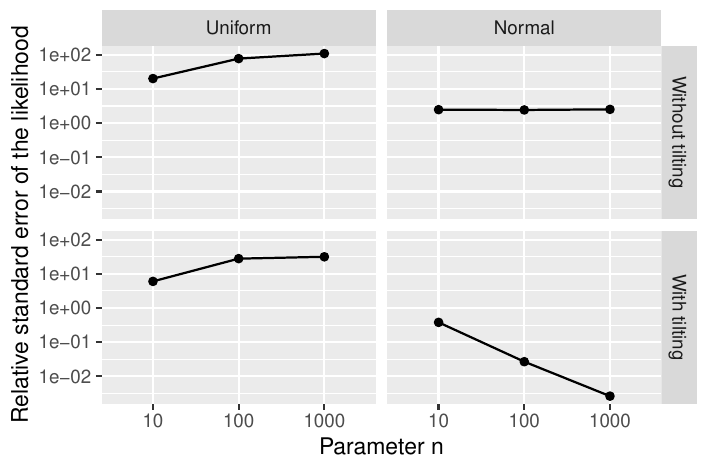}
    \caption{Comparison of the relative standard error (standard deviation
      divided by empirical mean, $10^3$ runs) of the 4 considered estimators
      with $N_{\text{sim}}$ = 10, as a function of $n$. See text for details on
    the experiment.}
	\label{fig:comparison_standard_error}
\end{figure}

\Cref{fig:comparison_standard_error} compares the relative standard error
(empirical standard deviation divided by empirical mean, over $10^3$
independent runs) of these four estimators in the following synthetic example:
we estimate the likelihood of  observations $y^1,\ldots, y^K$, where $y^k$ is a
realisation of $Y^k=AX^k$, $X^k\sim\multi(n, p)$,  for $k=1,\dots, K=100$,
$p=(1/9, \dots, 1/9)$,  $d_X=9$, and $A$ is the matrix described in
\cref{sub:formal} (for a two-round election with 3 candidates at each round).

We observe the following: both using the Gaussian proposal (rather than the
uniform) and using tilting reduces the variance. But it is the combination
of the two that really leads to a drastic variance reduction, as shown in the lower-right pane of \cref{fig:comparison_standard_error}. Furthermore, the
variance keeps decreasing as $n$ increases, in line with \cref{prop:cv}.

\Cref{sec:num} contains additional results on the performance of these
estimators in different synthetic scenarios. In particular, we find that the
uniform proposal may actually outperform the Gaussian proposal when $n$ is
small; see \cref{fig:uniform_vs_normal} and the surrounding discussion.  But,
for the type of applications we are considering in this paper (polling stations
where $n\gg 100$), this case is not relevant.

\section{Application to ecological inference}
\label{s:application}

We now apply saddlepoint Monte Carlo to ecological inference. Our particular examples
are taken from electoral sociology for elections with two rounds, for which
the analysis with traditional two-by-two EI schemes is impractical.

\subsection{Motivation}

We take the example of three important French elections of the last two
decades: the presidential election of 2007, the presidential election of 2022,
and the legislative elections of 2024 (to elect members of the French National
Assembly). All of these are elections in two rounds which exhibited non-trivial
vote carryover. In the French electoral system, any number of candidates may
take part in the first round. Only the top two vote-getters in the first round
qualify for the second round in presidential elections. In legislative
elections, between two and four candidates qualify for the second round.  Note
that voters have the option of abstaining from voting or of casting a blank or
spoilt ballot; we merge these options and handle this by adding "Abstention" to
the list of candidates at each round.

For the 2007 presidential election, Bayrou, a centrist candidate, was
eliminated with 18.6\% of the votes in the first round, and his voters had to choose between a
left-wing candidate (Royal) and a right-wing candidate (Sarkozy) in the second round. Bayrou did
not announce what electors should do nor what he would vote, and voice
carryover was crucial in the win of candidate Sarkozy.

For the 2022 presidential election, Mélenchon, a left/far-left
candidate, was eliminated with 19.6\% of the votes in the first round, and his voters had to
choose between a centrist candidate (Macron) and a far-right candidate (Le
Pen) in the second round. This situation was very similar the the 2017 presidential election,
except that commentators expected less voice carryover from Mélenchon's
electorate to Macron because of the increasing defiance towards his policies
and his figure. In particular, we will study how Mélenchon's electorate behaved
in the second round as a function of demographic density.

For the 2024 legislative elections, which took place in 577 constituencies, the
far right dominated the first round but was then defeated in the second round,
because of voice carryover in both directions between a centrist block and a leftist block, despite
the assumed weathering of the ``republican front'' against the French
far-right. The situation across
constituencies  was very
heterogeneous (from first-round voting and who was qualified for the second
round to second-round voting recommendations by candidates); we will demonstrate that our method can handle this heterogeneity.

For all three elections, we use data from the French Ministry of the Interior,
and in particular, data at the level of the around $60\, 000$ voting stations
in the country. For the second election, we merge this information with 
census data, which is publicly available by the national statistics bureau Insee at the level of townships. See
\cref{app:data} for more details on the data (obtained
\href{https://www.data.gouv.fr/fr/pages/donnees-des-elections/}{here}) and how
it was pre-processed before we carried out analysis.

Each of these studies aims to demonstrate one aspect of our method, from a
computational point of view. The 2007 presidential election allows us to
perform a general presentation of our inference scheme; to show that it scales
well to very large datasets; and that the multinomial model and the Gaussian model
are not equivalent. The 2022 presidential election allows us to show that our
scheme is very flexible with functional forms, for example studying marginal
effects of exogenous covariates, also showing that more flexible functional
forms can lead to an easier inference. The 2024 legislative elections allow us
to showcase how the inference procedure may scale with size (some
constituencies had many candidates), and for smaller datasets (with sometimes
only $60$ voting stations in one constituency), thus not in the asymptotic setting.

\subsection{Models}

We study elections in two rounds with $I$ options in the first round, $J$
options in the second round, and $K$ voting stations.  Let  $n^k$ be the number
of voters at station $k$ ($k\in\{1,\ldots,K\}$),  $\tilde{X}^k$ be  the 
$I\times J$ matrix which records the number of people who voted for candidate
$i$ in the first round and $j$ in the second round in station $k$, and let
$X^k=\vecop(\tilde{X}^k)$ be the vector obtained by stacking the columns of
$\tilde{X}^k$. We observe $y^k$, the realization of $Y^k = AX^k$, where $A$ is the
matrix such that $AX^k$ contains all the margins (sums over rows and over
columns) of the matrix $\tilde{X}^k$.  In words, in each station $k$, we
observe the total number of votes for each candidate at the first round, and
those at the second round. Technically, we drop the last row and column
marginals because they are redundant, as the analysis is conditional on $n^k$.
Thus, $d_X=I \times J$ and $d_Y=I+J-2$.

We call model 1 the natural baseline model
\begin{equation*}
  X^k \sim \mathcal{M}(n^k, p),\quad\text{for }k=1,\dots,K
\end{equation*}
where $p=\vecop(\tilde{p})$, $\tilde{p}=(\tilde{p}_{i,k})$ is a $I\times J$
matrix, 
\begin{equation*}
  p = \text{softmax} (0, \theta_1, ..., \theta_{IJ - 1}),
\end{equation*}
and the vector $\theta\in\R^{d_\theta}$, $d_\theta=IJ-1$, is assigned  prior
distribution $\pi(\theta) \sim \norm\left(0, \sigma^2 \mathrm{I}_{d_\theta}\right)$,
with $\sigma^2 = 2$. This prior distribution will only have a very minor effect
on the inference as the information provided by $60\, 000$ voting stations and more
than 45 million voters is high. However, this is preferable to a flat prior for
robustness purposes, as some probabilities may be very small, which, at the
extreme, can lead to a completely flat posterior density for certain
directions, which makes inference more difficult because of trivial options
selected by no voters.

It is easy to observe that candidates do not perform
homogeneously across constituencies, which makes our model misspecified. 
A simple fix that does not involve introducing a more complex model is to condition the analysis
on the first marginal, and only study conditional probabilities for the second
round (which are our core focus anyway). That is, parametrize the models in
terms of $p_{j|i}$ (the probability of voting $j$ at the second round, given
that one has voted for $i$ at the first round), rather that in terms of
$p_{ij}$. Our Model 2 thus assumes  $X^k\sim\multi(n^k, p^k)$, where $p^k=\vecop(\tilde{p}^k)$, 
\begin{equation*}
  \tilde{p}_{i,j}^k = \left( \frac{1}{n} \sum_{j'=1}^J\tilde{X}^k_{i,j'} \right)\times \tilde p_{j|i}
  \quad \forall i=1,\dots,I, j=1,\dots, J
\end{equation*}
and take finally
\begin{equation*}
\tilde p_{ \bdot|i} = \softmax(0, \theta_{i, 1}, ...,
  \theta_{i, J-1}).
\end{equation*}
The dimension of $\theta$ is then $d_\theta=(I-1)(J-1)$. 


Our framework allows to consider covariates.
For the 2022 presidential election, we will describe the effect of a single
continuous covariate at the level of the voting booth, 
$(C^k)_{1\leq k \leq K}$, and use a similar logistic linear regression
parametrization, this time leading to different conditional probabilities
across voting stations
\begin{equation*}
  \forall k, \forall i, \quad
  \tilde p_{.|i} = \text{softmax}(0, \theta_{i, 1} +  \beta_{i,1} C^k , ..., \theta_{i, J-1}+\beta_{i,J-1} C^k ),
\end{equation*}
with $\beta$ an array of parameters the same size of $\theta$, for which we use the same prior distribution $\mathcal{N}(0, \sigma^2 I)$, $\sigma^2= 2$.
We refer to this as Model 3.

\subsection{Computation}

We need to sample from the posterior distribution
\begin{align*}
  \pi(\theta \mid y_{1:K}) 
   & \propto \pi(\theta) \prod_{k=1}^K \pr_\theta(AX^k = y^k) \\
   & \propto \pi(\theta) \prod_{k=1}^K f_{AX_k}^\theta (y_k)
\end{align*}
where $f_{AX_k}^\theta$ is the density of $y_k$ under the considered model for
parameter $\theta$, as defined in the previous section.  Recall that
saddlepoint Monte Carlo makes it possible to obtain independent, unbiased estimators of
each factor of the above product, and therefore gives an unbiased estimator of the posterior density. On
the other hand, the log of the estimator of the posterior density is a biased
(although possibly useful) estimator of the log posterior density.

We find it most effective to proceed in two steps: first, derive
a Gaussian approximation of the posterior through an optimization scheme; 
second, use this Gaussian approximation to  calibrate a pseudo-marginal sampler
that is able to sample exactly from the posterior.

For the first step, we approximate the MAP (maximum a posteriori) estimator 
$\theta^\star = \arg\max_\theta \log \pi(\theta \mid y_{1:K})$, using  a variant of stochastic gradient descent (SGD) based
on minibatches. We then refine
this approximation using SGD but based on the whole dataset. More precisely, we
proceed as follows:
\begin{enumerate}
  \item Starting at $\theta = 0$, run $2\, 000$ iterations of
    Adam~\citep{kingma2014adam} with batches of size $\min({2\, 000, K/2})$, a
    learning rate of $10^{-1}$, yielding a first MAP approximation $\tilde
    \theta$. 

  \item Run $5\, 000$ iterations of Adam on the entire data with learning rate $10^{-2}$,
    initialized at $\tilde \theta$, with $N_{\text{Sim}}= 2^4$ for $4\, 500$ iterations
    and $N_{\text{Sim}} = 2^7$ for the last 500 iterations, and compute the average of
    coefficients over the last 50 iterations, yielding a better MAP
    approximation $\hat \theta^\star$. 

  \item Obtain an estimate $\hat{H}$ of the Hessian of the negative log-likelihood at $\hat
    \theta^*$. 
\end{enumerate}

The resulting Gaussian posterior approximation is then $\norm(\hat{\theta}^\star,
\hat{H}^{-1})$ (not to be confused with the Gaussian proposal within the
saddlepoint). Note that this first step is obviously approximate, since it
relies on derivatives (gradient, Hessian) of a biased estimate of the log
posterior density, as per the remark above.  These derivatives are obtained
through automatic differentiation.

In a second step, we use by default random weight importance sampling
\citep{randomweight}, with the
Gaussian approximation as a proposal. That is, we sample $\theta_n\sim
\norm(\hat{\theta}^\star, \hat{H}^{-1})$, $n=1,\ldots, N$, and assign to each
simulated value the random weight:
\begin{equation*}
  w_n = \frac{\hat{\pi}(\theta_n \mid y_{1:K})}{\varphi(\theta_n;
  \hat{\theta}^\star, \hat{H}^{-1})},
  \quad \hat{\pi}(\theta_n \mid y_{1:K}) = \pi(\theta_n) \prod_{k=1}^K
  \hat{f}_{AX^k}^\theta(y^k),
\end{equation*}
where $\hat{f}_{AX^k}^\theta(y^k)$ is a saddlepoint Monte Carlo estimate of 
$f_{AX^k}^\theta(y^k)$ (with tilting) and $\varphi(\bdot;\mu,\Sigma)$
denotes the probability density of a $\norm(\mu, \Sigma)$ distribution. 

This approach can indeed be referred to as pseudo-marginal, in the sense that
\begin{equation*}
  \E\left[w_n \mid \theta_n\right] = \frac{\pi(\theta|y_{1:K})}{\varphi(\theta_n;
  \hat{\theta}^\star, \hat{H}^{-1})},
\end{equation*}
the ideal importance sampling weights one would obtain if the posterior density $\pi(\theta \mid y_{1:K})$ were tractable. 
In particular, we have 
\begin{equation*}
  \frac{\sum_{n=1}^N w_n \psi(\theta_n)}{\sum_{n=1}^N w_n}
  \to \int \psi(\theta) \pi(\theta \mid y_{1:K}) \dx{\theta}
\end{equation*}
as $N\rightarrow \infty$ for a test function $\psi$, even if the number of
draws used within the saddlepoint Monte Carlo procedure remains constant.
See~\cite{randomweight} for more details on random weight importance sampling.

We find this approach to work well for datasets large enough to make the
posterior very close to Gaussian; this is the case in our results in
\cref{sub:president2007,sub:president2022}, where the data  corresponds to all
of France, or one of its regions. Alternatively, for a smaller dataset such as
a single constituency, as in  \cref{sub:legislatives}, one may use instead a
pseudo-marginal Metropolis sampler based on a random walk proposal (with the
covariance proportional to $\hat{H}^{-1}$). See~\cite{andrieu2009pseudo} for
more background on such pseudo-marginal samplers, which are able to sample from
the true posterior while having access only to unbiased estimates of its
density.

\subsection{Broad description of computing time} 

Obtaining the full posterior for one basic model over all of France's $60\,
000$ voting stations and an augmented dataset of more than one million lines
took around two hours, using one CPU on personal hardware, or around 5 minutes,
using one A100 GPU\footnote{Scripts and resulting data are available on the \href{https://github.com/theovoldoire/saddlepointMC-open/}{public repository}.}. On CPU, the first step took 3 minutes (minibatch Adam), plus
90 minutes (Adam on the whole data), while the second step (pseudo-marginal
sampling), around 30 minutes. 

The number of iterations at each step was selected conservatively and these
times could be reduced further; some experiments took under one hour of
computing time. In addition, this could be considerably sped up by using GPUs as
all operations are compatible with accelerated linear algebra (XLA) methods.
The time to obtain the posterior distribution for a single constituency of the
legislative elections with one CPU ranged, in the same conditions, between 5
and 10 minutes.

\subsection{Results: 2007 presidential election}\label{sub:president2007}

\subsubsection{Evaluating the approximate Gaussian model} 

Recall that we stressed how previous works in the EI literature have often
simplified the problem by replacing the multinomial distribution of $X$ with
some other (continuous) distribution, notably working with frequencies instead of
counts, under the justification of the central limit theorem.  In this section,
we evaluate the quality of this simplification. To this end, we  evaluate
whether our baseline model (model 1) $X^k \sim \mathcal{M}({n^k, p})$ and its
Gaussian approximation yield similar posterior distributions. We consider the
approximate Gaussian model proposed by \cite{wakefield2001ecological}:
\begin{equation*}
  X^k \sim \mathcal{N}\left(n^k p, n^k \left( \text{diag}(p)- p p^\top
  \right)\right).
\end{equation*}

\begin{figure}[h!]
	\includegraphics{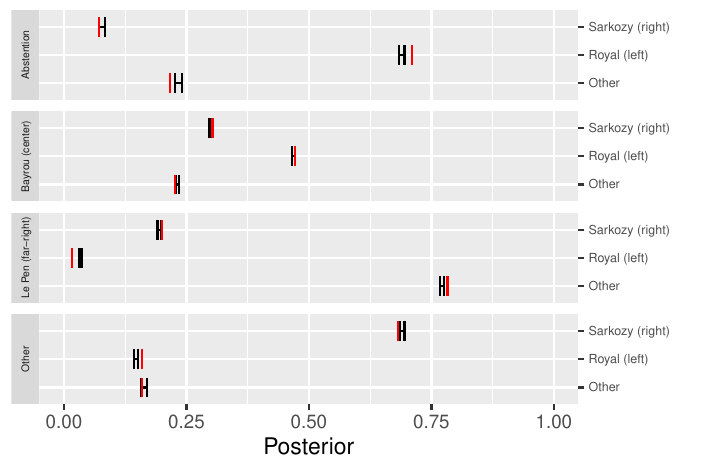}
    \caption{Comparison of the 90\% posterior interval for predicted
      probabilities of the true model 1 (black) and median posterior predicted
      for the approximate Gaussian model (red), conditional on first-round
      choices (right x-axis) and the different second-round choices (left
      x-axis). Decisions conditional on having voted Royal or Sarkozy in the
      first round are not represented here. Interpretation: most median
      predicted probabilties for the approximate model fall outside the 90\%
      interval of the true model.}
	\label{fig:comparison_gaussian}
\end{figure}


We present here a comparison restricted to the Ile-de-France region. 
The model includes $18$ probabilities to infer. Of these, $6$ are trivial (e.g., the probability of voting for Sarkozy in the second round conditional on having voted for Royal in the first round is essentially 0).
We compare the posterior distributions of the remaining $12$ parameters, obtained under the multinomial model and under the normal model; they are summarized  in \cref{fig:comparison_gaussian}. 
For ease of visualization, we display the 90\% credible interval under the true, multinomial model, and only the posterior median under the approximate, Gaussian model. 
We find that $9$ of the posterior median probabilities for the normal model do not
fall in the 90\% credible interval for the binomial model,
indicating that there is a substantial difference in the inference between the two approaches.
Examples of these discrepancies concern key quantities of interest, such as
the transition rates from Bayrou to Royal or from Le Pen to Abstention.
It is worth pointing out that there exists in the literature a misconstrued belief that the Gaussian approximation should at least be appropriate for conditions with large marginal counts; this example shows that this belief is erroneous, as seen for example with the badly inferred probability of switching from Abstention to Royal: this concerns a large count of individuals, but the approximate posterior does not match the true posterior.

This simple example shows that the Gaussian
approximation does not necessarily approximate well the binomial distribution
and that researchers interested in ecological inference should probably be more
cautious when substituting the analysis of counts in a voting station with an
analysis of frequencies.

\subsubsection{Analysis of the outcome}

We now turn to model 2 and provide methodological and substantive conclusions.
We perform inference on all $60\, 000$ French voting stations. We  sample $6\,000$ points
from the Laplace approximation; we obtain an ESS of 280, which suggests that,
despite being of the pseudo-marginal family, our method scales fairly well to
large datasets. Furthermore, the standard error of the log-likelihood is
estimated at $1.60$, which   is very encouraging given
the size of the application which concerns over 40 million French voters.
Results are presented in \cref{tab:full_model_2007}.

Substantively, this model is interesting as it is different from the study
performed in some exit polls. For example, in Ipsos' exit poll (on 200 voting
stations, whereas ours is exhaustive), 25\% of Le Pen voters and 21\% of Bayrou
voters were estimated to have voted blank or to have abstained, where our
estimated proportions are only of 16\% for both. One interpretation is that
some individuals did not disclose their real vote in the first round, putting
more weight on voters committed to not choose between the two remaining
candidates; another is that individuals do not want to announce that they voted
for the two candidates and lie that they were keen on refusing this binary
choice. 
Among people who had abstained in the first round, the Ipsos exit poll estimated that only 64\% abstained in the second round, which is far
 from our estimate of 80\%; this could also be explained by desirability bias in the
exit polls.

\begin{table}
	\centering
	\begin{tabular}{lrrr}
\toprule
 & Sarkozy & Royal & Other \\
\midrule
Sarkozy & 0.97 & 0.00 & 0.03 \\
Royal & 0.00 & 0.98 & 0.02 \\
Bayrou & 0.49 & 0.36 & 0.15 \\
Le Pen & 0.71 & 0.13 & 0.16 \\
Abstention & 0.07 & 0.14 & 0.80 \\
Other & 0.25 & 0.63 & 0.11 \\
\bottomrule
\end{tabular}

	\caption{Estimated transition rates between the first and second round of the 2007 presidential election across all voting precincts for model 2. Only the median of the posterior is presented as the posterior is extremely concentrated (posterior intervals at the 90\% level are at most of length $0.01$.)} \label{tab:full_model_2007}
\end{table}

\subsection{Results: 2022 presidential elections}\label{sub:president2022}

Having demonstrated that our method can scale well to large datasets, we now
turn to the 2022 presidential election, where we will exemplify that our
method is compatible with more flexible model forms, and in particular with
covariates (using models 2 and 3). We focus on the
second-round voting behavior of first-round Mélenchon (left/far-left) voters. We first focus on
Ile-de-France and compare second-round voting across departments, which enables
a broad geographical comparison. We then turn back to all of France and
consider model 3, where the relationship of voting with population density is
considered.

\subsubsection{Geographical comparison}

\begin{figure}[h!]
	\includegraphics{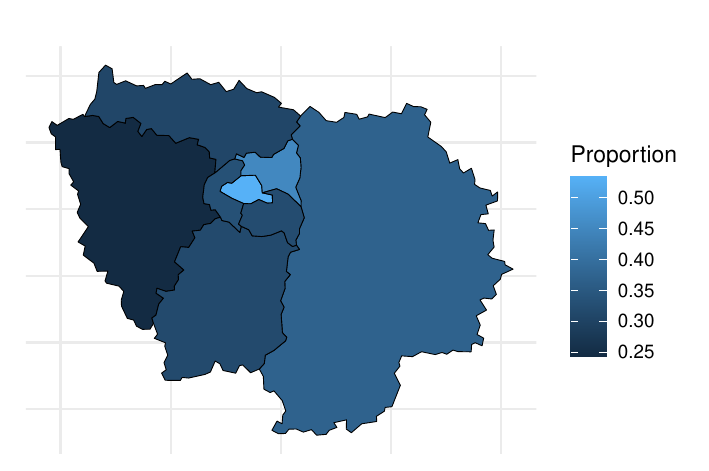}
	\caption{Median estimated proportion of voters who voted in the second round for Macron (centre) after voting for Mélenchon (left/far-left) in the first round, estimating separately model 2 for each department in Ile-de-France. Almost all remaining voters either abstained or voted a spoilt ballot.}
	\label{fig:comparison_depts}
\end{figure}

In Ile-de-France, we infer that less than 1\% of Mélenchon voters decided to vote for Le Pen (far-right) in the second round, in all departments.
The quantity of interest is thus whether Mélenchon voters carry over to Macron
or to Abstention in the second round. We show on  \cref{fig:comparison_depts}
the proportion of Mélenchon voters who carried over to voting for Macron
(centre) in the second round. 
As earlier, the posterior is quite concentrated and we only report  the
posterior median in the figure.

We find that the
proportion of people who voted for Macron after voting for Mélenchon is highest
in Paris and Seine-Saint-Denis, at 53.4\% and 45.2\%. We find in an
intermediate position several departments: Seine-et-Marne (37.1\%),
Hauts-de-Seine (33.2\%), Val-de-Marne (32.2\%), Essone (31.7\%) and Val d'Oise
(30.9\%). Finally, one department stands out as having a higher proportion of
people who did not vote for Macron in the second round after voting for
Mélenchon, with Yvelines, at 24.3\%. We find this set of observations quite in
line with what could be expected, with more anti-far-right voting in Paris and
Seine-Saint-Denis. These results show strong geographical heterogeneity in voting patterns, and indicate that a more complex model may be necessary. Indeed,
a motivation for doing ecological inference is to
adopt more fine-grained tools and variables than broad geographical ensembles,
and we now move to study the effect of a continuous variable: population
density.

\subsubsection{Model comparison for population density}

To study the effect of population density, we estimate models 2 and 3 on all
of France. We draw $3\,000$ values from the Laplace approximation, giving an
ESS of 551 for model 2 (fixed probability) and of $1\,074$ for model 3 (with
density covariate). A possible explanation is that the standard error of the
estimator of the log-likelihood is slightly smaller at the MLE for model 3 than
for model 2 and is an example of why we hypothesize inference may be easier for
better-specified models even if the parameter count is larger. We compute the
Bayes factor $\mathrm{BF}_{3 / 2}$ by approximating the marginal likelihood for each model
using random weight importance sampling with the Laplace approximation as the
proposal distribution, which provides more than decisive evidence in favor of model 3,
with 
\begin{equation*}
  \log_{10} \mathrm{BF}_{3 / 2} = 62\,058 > 2.
\end{equation*}
This is another interest of marginal likelihood methods, namely that they
provide access to the marginal likelihood and so enable quick and easy
comparison across models.

\subsubsection{Analysis of posterior probabilities}
\begin{figure}[h!]
	\includegraphics{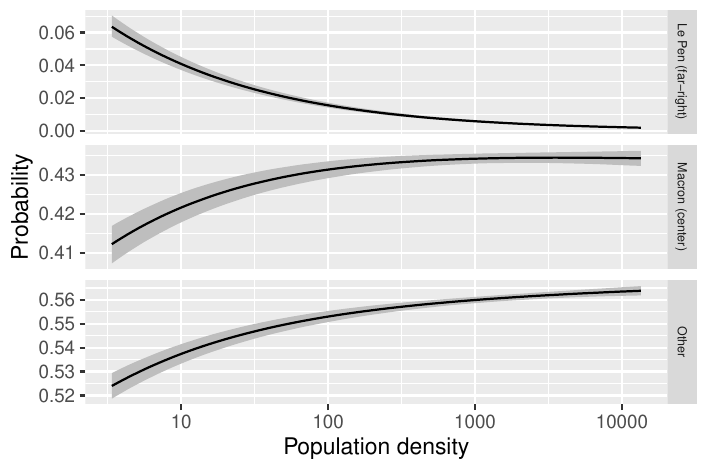}
    \caption{Predicted probabilities for second round behavior after voting
      Mélenchon (left/far-left) by population density of the city in which the
      voting station is located. Ribbons represent the $[.5, .95]$ posterior
    intervals.}
	\label{fig:density}
\end{figure}

Now that it is clear that one should prefer the model with population density,
we focus on the specific case of Mélenchon-to-second round voting. Again, for
simplicity of interpretation, we transform the posterior draws of coefficients
into posterior predicted probabilities, this time over a grid of population
densities. Results are presented in \cref{fig:density}.

We observe that among first round Mélenchon voters,  the probability to
vote for Macron in the second round increases with the population density of the
city in which the voting station is located. However, we also find
that abstention also becomes more likely as the
density increases and that the probability for voting far-right drastically
decreases, from around 6\% when the density is lowest, to near 0\% when the
density is highest. This decrease of six percentage points is spread across
voting Macron (2 percentage points) and abstaining  (4
percentage points). A possible interpretation is that electors in more dense
areas listened more to the recommendation of Mélenchon after the first round,
which was that ``no vote should go for the far-right'', but that this
recommendation was less followed by individuals in less-populated areas.

\subsection{2024 legislative elections}\label{sub:legislatives}

We now consider the 2024 snap legislative elections. We focus in particular on
the so-called ``front républicain'' (republican front) across centrist and leftist
electorates, which corresponds to strategic considerations to attempt to avoid a far-right win. 
We consider all constituencies where the top three candidates in the first round are one each from the far-right, the centre, and the (far-)left. In all the constituencies we examine, the far-right candidate qualified for the second round; depending on the specific cases, either one or both of the other candidates also qualified. Note that electoral law allows for more than two candidates to qualify for the second round in certain cases; when three candidates qualify, this is referred to as a \emph{triangular} second round. 
The umbrella term ``front républicain'' groups several  strategic considerations for centre and left candidates and voters wishing to avoid a far-right win in the second round.
First, in triangular second round elections, the weaker candidate between centre and left may withdraw from the second round and instruct their voters to vote for the stronger candidate; in such cases, we wish to evaluate the impact of this decision.
Second, the weaker candidate may decide to stay in the race, but their voters may nonetheless switch to the stronger candidate in the second round; here too, we wish to estimate how often this occurs.
Third, when there are only two candidates in the second round, the voters of the non-qualified candidates have to choose between abstaining, or voting for a candidate they dislike in order to avoid a far-right win.
This question is of interest to political scientists working on the proximity across other political systems than France which exhibit three blocks, at the
centre/centre-right, left/far-left, and far-right.

We excluded constituencies for French living abroad (they only have one
voting station and so  EI cannot be used) and were left with 312 constituencies.
We then split these constituencies according to two criteria: 
whether the left candidate arrived ahead of behind the centre candidate in the first round, and
whether the weaker of those two candidates remained in the second
round or not (be it because they did not qualify, or because they withdrew). This defines four
situations: left ahead and centre  out (122 constituencies); left ahead and
centre remains (48); centre ahead and left  out (121); and centre ahead and
left remains (21). Recall that in all 312 cases, the far-right candidate remained in the second round.

\subsubsection{Two constituency examples}

To be more concrete, we present the estimated coefficients in two politically
salient constituencies: Calvados 6 and Somme 1. In 
Calvados 6, Elisabeth Borne, a centrist who had recently resigned as prime minister,
received 28.9\% of the votes in the first round against far-right candidate Nicolas Calbrix
who received 36.3\%; the leftist candidate Noé Gauchard qualified for the second round but withdrew and Elisabeth Borne ultimately won with 56.4\% of the votes in the second
round. In Somme 1, François Ruffin, a high-profile and left/far-left
candidate, obtained 33.9\% of the votes in the first round, against far-right
candidate Nathalie Ribeiro-Billet  who obtained 41.7\%. The centrist candidate Albane Branlant qualified for the second round but withdrew, and François Ruffin ultimately won in the second round
with 52.9\% of the votes.

\begin{table}
	\tiny
	\centering
	\begin{tabular}{lrrrrrrrrrrrr}
\toprule
 & \multicolumn{3}{r}{abstention} & \multicolumn{3}{r}{CALBRIX RN} & \multicolumn{3}{r}{BORNE ENS} & \multicolumn{3}{r}{other} \\
 & .05 & .5 & .95 & .05 & .5 & .95 & .05 & .5 & .95 & .05 & .5 & .95 \\
\midrule
abstention & 0.71 & 0.73 & 0.75 & 0.14 & 0.16 & 0.18 & 0.05 & 0.07 & 0.08 & 0.03 & 0.04 & 0.05 \\
CALBRIX RN & 0.08 & 0.10 & 0.12 & 0.84 & 0.87 & 0.89 & 0.00 & 0.01 & 0.02 & 0.01 & 0.02 & 0.03 \\
GAUCHARD UG & 0.07 & 0.09 & 0.11 & 0.01 & 0.01 & 0.03 & 0.80 & 0.82 & 0.84 & 0.06 & 0.08 & 0.10 \\
LAHALLE DVC & 0.12 & 0.18 & 0.24 & 0.04 & 0.08 & 0.14 & 0.57 & 0.64 & 0.70 & 0.05 & 0.10 & 0.17 \\
BORNE ENS & 0.05 & 0.07 & 0.09 & 0.01 & 0.03 & 0.05 & 0.86 & 0.88 & 0.90 & 0.01 & 0.02 & 0.03 \\
other & 0.21 & 0.29 & 0.37 & 0.19 & 0.26 & 0.35 & 0.14 & 0.21 & 0.28 & 0.18 & 0.24 & 0.30 \\
\bottomrule
\end{tabular}

	\begin{tabular}{lrrrrrrrrrrrr}
\toprule
 & \multicolumn{3}{r}{abstention} & \multicolumn{3}{r}{RIBEIRO B. RN} & \multicolumn{3}{r}{RUFFIN UG} & \multicolumn{3}{r}{other} \\
 & .05 & .5 & .95 & .05 & .5 & .95 & .05 & .5 & .95 & .05 & .5 & .95 \\
\midrule
abstention & 0.88 & 0.89 & 0.90 & 0.03 & 0.03 & 0.04 & 0.07 & 0.08 & 0.09 & 0.00 & 0.00 & 0.01 \\
RIBEIRO B. RN & 0.03 & 0.04 & 0.05 & 0.93 & 0.94 & 0.95 & 0.00 & 0.01 & 0.01 & 0.00 & 0.01 & 0.01 \\
BRANLANT ENS & 0.03 & 0.05 & 0.07 & 0.16 & 0.17 & 0.19 & 0.51 & 0.53 & 0.56 & 0.23 & 0.24 & 0.26 \\
RUFFIN UG & 0.03 & 0.04 & 0.06 & 0.00 & 0.00 & 0.01 & 0.94 & 0.95 & 0.97 & 0.00 & 0.00 & 0.01 \\
other & 0.08 & 0.13 & 0.20 & 0.18 & 0.26 & 0.34 & 0.41 & 0.51 & 0.60 & 0.04 & 0.10 & 0.16 \\
\bottomrule
\end{tabular}

	\caption{\footnotesize{Secound round behavior in Calvados 6 and  Somme 1. Probabilities of voting in the second round (columns) conditional on first-round behavior (rows), with median posterior probability and 90\% credibility intervals. Political labels meaning: RN: far-right, UG: left, ENS \& DVC: centre.}}
	\label{fig:borne_ruffin}
\end{table}

Results for these cases are available in \cref{fig:borne_ruffin}. In Calvados 6, among voters who cast their vote for the leftist candidate Gauchard in the first round, around  82\%  then voted for Elisabeth Borne in the second round.
In Somme 1, among voters who cast their vote for centrist candidate Branlant in the first round,  around 53\%  then voted for
François Ruffin in the second round. Both these constituencies showcase rather
high conversion rates in their respective categories (left-to-centre and
centre-to-left). On the other hand, two other constituencies exhibited, when
compared to their own category, lower conversion rates for national political
figures, and are available in the appendix, with Gérald Darmanin (66\% of
left-to-centre conversion) and Antoine Léaument (22\% of centre-to-left
conversion).

\subsubsection{Analysis of all constituencies}

\begin{figure}
	\centering
	\includegraphics{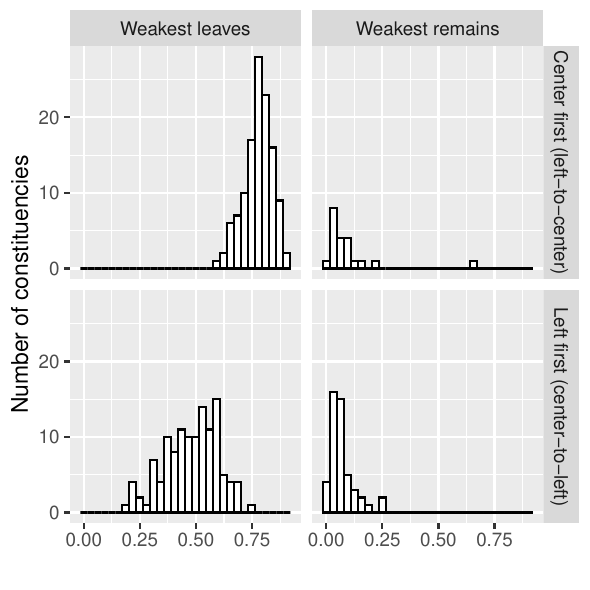}
	\caption{\footnotesize{Histogram of the median predicted probability across all constituencies, in the four situations described. Upper panes correspond to left-to-centre probabilities and lower panes to centre-to-left.}}
	\label{fig:outcomes_all_constituencies}
\end{figure}

We now analyse the output on all 312 constituencies we considered; the results are summarized in \cref{fig:outcomes_all_constituencies}.
We first examine the impact of the presence of a third candidate in the second round (right panels in \cref{fig:outcomes_all_constituencies}) vs cases where the weaker candidate either did not qualify or withdrew (left panels in \cref{fig:outcomes_all_constituencies}).
As expected, there is a much lower conversion rate in cases where the weakest-performing
candidate remains in the race for the second round. 
When
the weaker candidate (between centre and left) was not present in the second
round, the average of median probabilities is much higher for left-to-centre
cases (at 0.778) than for centre-to-left cases (at 0.480). Note also that there is a
much higher heterogeneity when considering the centre-to-left cases, with a
standard deviation of 0.119 against 0.062 for left-to-centre cases. We 
evaluate below how this relates to the differences across the different parties that
were part of the left movement for the election.


 We also observe that even when the weakest candidate remains in the second round, some of their electors voted in
the second round for the stronger candidate (right panes in \cref{fig:outcomes_all_constituencies}). The transition rates to the strongest candidate, both for
left-to-centre and for centre-to-left, are much smaller, but not equal to zero,
and sometimes as high as 0.25. The outlier in the upper-right pane (at 0.62)
is a candidate who wished to withdraw between the two rounds but failed to produce the proper paperwork on time. This analysis suggests that candidates in a position to withdraw
can be a major driver in the second-round behavior of electors, but one should not
assume that transition rates will be close to 1, especially for centre-to-left situations.

\subsubsection{Impact of two covariates}
\begin{figure}
	\hspace*{-1cm}
	\centering
	\includegraphics{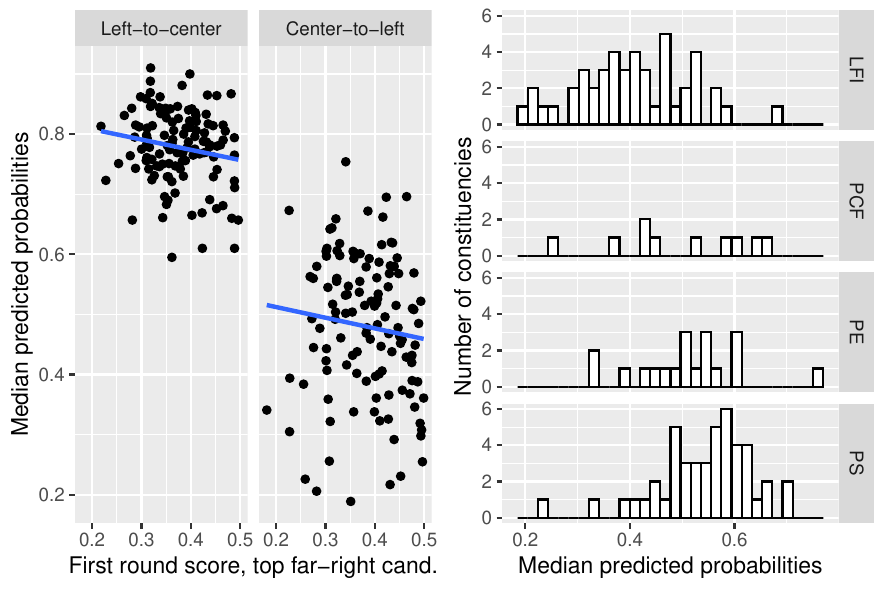}
	\caption{Left pane: relationship between the first round score of the top far-right candidate and the probability of voting for the remaining candidate; each dot corresponds to a constituency in the 2024 legislative elections. The blue lines are the least-squares regression lines.
		\\Right pane: distribution for centre-to-left median transition probabilities depending on the which left party is present.}
	\label{fig:other_hypotheses}
\end{figure}

We now assess two additional hypotheses. First, are electors voting more
strategically when the far-right candidate obtained a higher score in the first
round? Second, is centre-to-left voting behavior the same depending on
the type of left candidate present in the constituency, either from the
traditional left (socialist party, PS), the communist left (communist party,
PCF), the ecologist left (the ecologists, PE), or the populist left
(``unbowed France'', LFI)? We now include these covariates in the analysis; results are summarized
in \cref{fig:other_hypotheses}.

The left pane of \cref{fig:other_hypotheses} shows that the assumption
of a relationship between cross-over voting and far-right scoring seems false. Centre-to-left and
left-to-centre transition rates seems either to stay constant or decrease as
the share of the top far-right candidate in the first round increases. 
The small negative relationship may indicate that constituencies that lean more
on the far-right are in a different ideological space, in the sense that the
left is not seen as a credible or safe alternative, even for centrist voters. 

The right pane of \cref{fig:other_hypotheses} shows that
the centre-to-left transition rate is substantially higher when the left
candidate hails from PS rather than from LFI (the number of constituencies is
too low to conclude for left candidates from PCF and PE).  The average for PS
is $0.510$ (sd: $.092$ for $46$ constituencies) where it is only $0.406$ (sd:
$0.108$ for 46 constituencies) for LFI. Note however that the difference
between these cases is small compared to  the difference with left-to-centre
transition rates.

\section{Future work}\label{sec:conclusion}

We have focused in this paper on ecological inference, as this is already  an
important class of problems for applied scientists.  We believe however that
saddlepoint Monte Carlo has a very promising potential in various other areas,
such as a data privacy  (where only aggregates are reported, e.g., quantiles,
to protect the privacy of individuals), or ill-posed inverse problems in
machine learning (i.e., when ones observes a noisy of exact version of $Y=AX$),
in a general sense. This also means pushing the method to its limits when $d_X$
or $d_Y$ get very large, a question we have not yet explored. 

As a simple example of a potential application, consider the ecological
inference model considered in this paper (that is, with the same type of matrix
$A$ that computes row and column margins), but with the components of $X$ being
independent Bernoulli$(1/2)$ variables. Then $\pr(AX=y)$ will be equal to the
number of binary matrices respecting the row and column constraints given by
$y$,  divided by the total number of such matrices.  In other words, we can use
saddlepoint Monte Carlo to approximate the number of contingency tables with
fixed margins, which appear in certain non-asymptotic tests; or alternatively
the number of Latin squares of a given order; this approach may be an
alternative to the SMC sampler of~\cite{chen2005sequential}.

\bibliography{main}

\appendix

\section{randomised quasi-Monte Carlo}\label{app:rqmc}

A RQMC (randomised quasi-Monte Carlo) sequence is a collection of $N$ random
variables, $U_1,\dots, U_{N}$, such that (a) each variable  $U_n\sim
\mathcal{U}[0,1]$ marginally, and (b) with probability one, the $N$ variables
$U_1,\dots, U_{N}$ have low discrepancy, that is, their star discrepancy is
$\mathcal{O}(N^{\varepsilon-1})$ for any $\varepsilon>0$. Whenever it is
possible to express an estimator as a deterministic function of $N$
independent $\mathcal{U}[0,1]$ variables, one may obtain a second estimator,
with the same expectation and (typically) much lower variance, by replacing
these independent variates with an RQMC sequence.  In our case, this is easy to
do, either for the uniform proposal (where the samples are already uniformly
distributed) or for the Gaussian proposal, where one can use the standard
inverse CDF trick, that is, for $Z_n\sim \norm(0, \Sigma_Y^{-1})$, take $Z_n =
C V_n$, where $C$ is the lower Cholesky triangle of $\Sigma_Y^{-1}$, $V_n$
a vector with components $V_{n}^j = \Phi^{-1}(U_{n}^j)$, and $\Phi$ is the CDF
of a $\norm(0, 1)$ distribution.

For more background on RQMC, again we refer to the books of~\cite{book_lemieux}
and~\cite{practicalqmc}. 

\section{Adapting saddlepoint Monte Carlo to a continuous distribution}\label{app:continuous_case}

In case $X$ takes values in $\R^{d_X}$, and thus $Y=AX$ takes values in
$\R^{d_Y}$, the likelihood \cref{eq:density_AX} is then the probability density
of $Y$, and it may  be expressed as:
\begin{equation*}
  f_{AX}(y) =
  \frac{1}{(2\pi)^{d_Y}} \int_{\R^{d_Y}} \exp\left\{-iz^\top y\right\} \varphi_X(A^\top z) \dx{z}
\end{equation*}
i.e. the only modification is the domain of integration. The same modification
must be applied to~\cref{eq:is_identity_for_AX}.

One could consider more generally the case where $X$ have both discrete and
continuous components, by changing the domain accordingly.

\section{Proof of \cref{prop:cv}}\label{app:proof_convergence}

Assume $X=X_{n}\sim\multi(n,p)$. Consider arbitrary (for now) sequences
$(y_{n})$, $(\nu_{n})$ and $(\mu_{n})$ in $\Z^{d_{Y}}$, and let
$\rho_{n}\coloneqq A^{\top}\nu_{n}$, $V_{n}\coloneqq\sqrt{n}(AX_{\rho_{n}}/n-\mu_{n})$.
Then we have:
\begin{align}
f_{AX_{n}}(y_{n}) & =\frac{M_{X}(A^{\top}\nu_{n})}{\exp(\nu_{n}^{\top}y_{n})}\int_{[-\pi,\pi]^{d_{Y}}}\exp(-iz^{\top}y_{n})\varphi_{AX_{\rho_{n}}}(z)\mathrm{d}z\nonumber \\
 & =\frac{M_{X}(A^{\top}\nu_{n})}{\exp(\nu_{n}^{\top}y_{n})}\int_{[-\pi,\pi]^{d_{Y}}}\exp\left\{ -iz^{\top}\left(y_{n}-n\mu_{n}\right)\right\} \varphi_{V_{n}}(\sqrt{n}z)\mathrm{d}z\nonumber \\
 & =\frac{M_{X}(A^{\top}\nu_{n})}{\left(\sqrt{n}\right)^{d_{Y}}\exp(\nu_{n}^{\top}y_{n})}\int_{[-\sqrt{n}\pi,\sqrt{n}\pi]^{d_{Y}}}\exp\left[-iz^{\top}\left\{ \sqrt{n}\left(\frac{y_{n}}{n}-\mu_{n}\right)\right\} \right]\varphi_{V_{n}}(z)\mathrm{d}z.\label{eq:tilted_dens}
\end{align}

Let $\left(\Sigma_{n}\right)$ be an arbitrary (for now) sequence
of invertible $d_{Y}\times d_{Y}$ matrices, $\varphi_{n}$ the characteristic
function of $\norm(0,\Sigma_{n})$, and $\psi_{n}$ the probability
density function of distribution $\norm(0,\Sigma_{n}^{-1})$. (Note
how the matrix is inverted in in the latter case but not in the former).
We have
\[
\psi_{n}(z)=c_{n}\varphi_{n}(z),\quad\text{with }\varphi_{n}(z)=\exp\left(-\frac{1}{2}z^{\top}\Sigma_{n}z\right),\quad c_{n}=\frac{\left|\Sigma_{n}\right|^{1/2}}{(2\pi)^{d_{Y}/2}}.
\]
We can then rewrite \cref{eq:tilted_dens} as:
\[
f_{AX_{n}}(y_{n})=\frac{M_{X}(A^{\top}\nu_{n})}{c_{n}\left(\sqrt{n}\right)^{d_{Y}}\exp(\nu_{n}^{\top}y_{n})}\int_{[-\sqrt{n}\pi,\sqrt{n}\pi]^{d_{Y}}}\exp\left[-iz^{\top}\left\{ \sqrt{n}\left(\frac{y_{n}}{n}-\mu_{n}\right)\right\} \right]\frac{\varphi_{V_{n}}(z)}{\varphi_{n}(z)}\psi_{n}(z)\mathrm{d}z.
\]
Consider an importance sampling (IS) estimate of this density, based
on proposal $\psi_{n}:$ 
\[
\hat{f}_{AX_{n}}(y_{n})=\frac{M_{X}(A^{\top}\nu_{n})}{c_{n}\left(\sqrt{n}\right)^{d_{Y}}\exp(\nu^{\top}y_{n})}\times\frac{1}{N_{\mathrm{IS}}}\sum_{m=1}^{N_{\mathrm{IS}}}\mathrm{Re}\left[\exp\left[-iZ_{m}^{\top}\left\{ \sqrt{n}\left(\frac{y_{n}}{n}-\mu_{n}\right)\right\} \right]\frac{\varphi_{V_{n}}(Z_{m})}{\varphi_{n}(Z_{m})}\right]
\]
where $Z_{m}\sim\norm(0,\Sigma_{n}^{-1})$. Its relative variance
is then: 
\[
\var\left[\frac{\hat{f}_{AX_{n}}(y_{n})}{f_{AX_{n}}(y_{n})}\right]=\frac{1}{N_{\mathrm{sim}}}\times\frac{\var\left[\mathrm{Re}\left\{ \exp\left[-iZ_{}^{\top}\left\{ \sqrt{n}\left(\frac{y_{n}}{n}-\mu_{n}\right)\right\} \right]\frac{\varphi_{V_{n}}(Z)}{\varphi_{n}(Z)}\right\} \right]}{\left(\E\left[\mathrm{Re}\left\{ \exp\left[-iZ_{}^{\top}\left\{ \sqrt{n}\left(\frac{y_{n}}{n}-\mu_{n}\right)\right\} \right]\frac{\varphi_{V_{n}}(Z)}{\varphi_{n}(Z)}\right\} \right]\right)^{2}}
\]
 with $Z\sim\norm\left(0,\Sigma_{n}^{-1}\right)$.

Recall that, when $X_{n}\sim\multi(n,p)$, the tilted variable is $X_{\rho_{n}}\sim\multi(n,q_{n})$
with $q_{n}=\xi(\rho_{n})$, where $\xi$ is defined above. Our tilting
strategy is to set $\eta_{n}$ such that $\E[AX_{\rho_{n}}]=nAq_{n}=y_{n}$,
that is $\eta_{n}=\lambda^{-1}(y_{n}/n)$, where function $\lambda$
was defined above (and we assumed it was a bijection, at least locally
around $t$). If we also set $\mu_{n}=y_{n}/n$, then we get 
\[
\var\left[\frac{\hat{f}_{AX_{n}}(y_{n})}{f_{AX_{n}}(y_{n})}\right]=\frac{1}{N_{\mathrm{IS}}}\frac{\var\left[\mathrm{Re}\left\{ \frac{\varphi_{V_{n}}(Z)}{\varphi_{n}(Z)}\right\} \right]}{\left(\E\left[\mathrm{Re}\left\{ \frac{\varphi_{V_{n}}(Z)}{\varphi_{n}(Z)}\right\} \right]\right)^{2}}
\]
and we can obtain our result by showing that $\varphi_{V_{n}}$
and $\varphi_{n}$ converges point-wise to same limit $\varphi_{\infty}$
, and thus $\varphi_{V_{n}}/\varphi_{n}\to1$, and applying the dominated
convergence theorem (since characteristic functions are bounded).
We assume that $\Sigma_{n}\rightarrow\Sigma$ deterministically, for
a certain matrix $\Sigma$, so that $\varphi_{n}\rightarrow\varphi_{\infty}$,
where $\varphi_{\infty}$ is the characteristic function of $\mathcal N(0,\Sigma)$.
What's left to prove is that $\varphi_{V_{n}}\to\varphi_{\infty}$
(and to choose a certain $\Sigma$). 

To that aim, we show that there exists $\Sigma$ such that
\begin{equation}
V_{n}=\sqrt{n}\left(AX_{\rho_{n}}/n-\mu_{n}\right)\rightarrow N\left(0,\Sigma\right)\label{eq:Vn}
\end{equation}
in distribution as $n\rightarrow\infty$. This implies that $\varphi_{V_{n}}\to\varphi_{\infty}$
point-wise, by Lévy's continuity theorem. To show~\cref{eq:Vn},
we specialise our results to $y_{n}=\lfloor nt\rfloor$, hence $\mu_{n}=\lfloor
nt\rfloor/n\to t$ as $n\to\infty$. We decompose $V_{n}$ as follows:
\begin{equation}\label{eq:decompo}
V_{n} = \sqrt{n}\left(A\tilde{X}_{n}/n-t\right)
        +\sqrt{n}\left(AX_{\rho_{n}}/n-A\tilde{X}_{n}/n\right)
        +\frac{1}{\sqrt{n}}(t-\mu_{n})
\end{equation}
where $\tilde{X}_{n}$ is such that (a)$\tilde{X}_{n}\sim\multi(n,q_{t})$,
for $q_{t}=\xi(A^{T}\eta_{t})$, and $\eta_{t}=\psi^{-1}(t)$; and
(b) $\tilde{X}_{n}$ is coupled with $X_{\rho_{n}}$ so that $\Vert AX_{\rho_{n}}/n-A\tilde{X}_{n}/n\Vert=\mathcal{O}_{P}\left(\|Aq_{n}-Aq_{t}\|\right)$.
We can obtain such a coupling by introducting $U_{1},\ldots,U_{n}$ independent
$\mathcal{U}[0,1]$ variables, and defining
\begin{align*}
X_{\rho_{n},i} & =\#\left\{ k:\,1\leq k\leq n,\ \sum_{j=1}^{i-1}q_{n,j}\leq U_{k}\leq\sum_{j=1}^{i}q_{n,j}\right\} ,\\
\tilde{X}_{n,i} & =\#\left\{ k:\,1\leq k\leq n,\ \sum_{j=1}^{i-1}q_{t,j}\leq U_{k}\leq\sum_{j=1}^{i}q_{t,j}\right\} .
\end{align*}

In the decomposition~\cref{eq:decompo}, it is obvious that the third term
converges to zero, as $\mu_{n}=\lfloor tn\rfloor/n$. This is also true for the
second term, since $Aq_{n}-Aq_{t}=y_{n}/n-t=\mathcal{O}_{P}(n^{-1}).$ 

Finally, for the first term, we apply the standard central limit theorem.
Thus, we need to set $\Sigma=AF^{-1}A^{\top}$ , where $F$ is the
Fisher information of the multinomial model at $q_{t}$. 

\section{Extra numerical experiments on synthetic data}\label{sec:num}

\subsection{Variance of the log-likelihood given $n$}

\begin{figure}[h!]
	\includegraphics{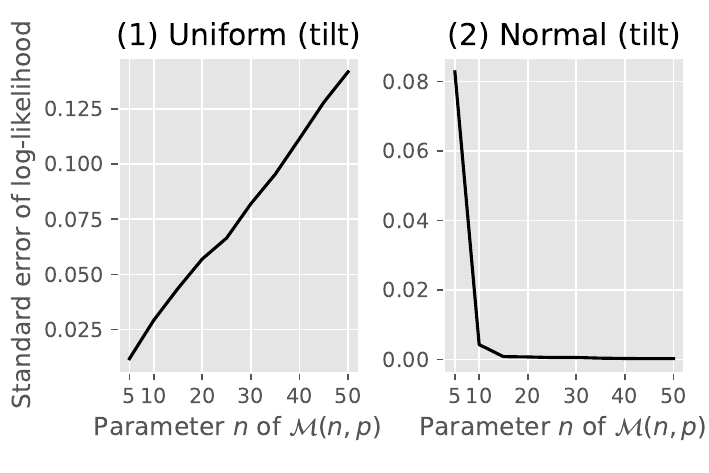}
    \caption{Standard error of the log of the estimated likelihood when 
      using either the uniform proposal (left) or the Gaussian proposal (right)
    in conjunction with tilting.}
	\label{fig:uniform_vs_normal}
\end{figure}

We consider the same settings as in \cref{ss:summary}, except we take $K=1$
(one observation $y_k$), and consider a range of smaller values for $n$. 
We compare the two proposal distributions (uniform vs Gaussian)
for the tilted estimators, in terms of the variance of the log-likelihood this
time. (For the experiment in \cref{fig:uniform_vs_normal}, we considered the
relative variance of the likelihood instead, because the non-titled estimators
occasionally returned negative values.)

Results are presented in \cref{fig:uniform_vs_normal}. They exhibit two
drastically opposed behaviors. The uniform sampling scheme, even with tilting,
has a standard deviation that linearly increases with parameter $n$. As $n$
increases, more and more independent variables $\mathcal{M}(1, p)$ are
considered, and the standard deviation of the logarithm of the likelihood of
their sum increases proportionally. In addition, except for $n = 5$, the
standard deviation is too high  to be used
in an actual application, even with $N_{\text{IS}} = 20\,000$. With a standard deviation of $0.05$, the strategy
would only work for a dataset composed of $20$ units or less. On the other
hand, the normal tilted estimator has a much lower variance, which drastically
decreases when $n$ increases. Except when $n = 5$ where the standard deviation
explodes, it is very close to zero. For $n = 50$, it is of $2.9 \times
10^{-4}$. Not shown in the figure, it is of $1.3 \times 10^{-5}$ for $n =1000$,
which is the size of many voting units we will study in the real data
application. Such small errors indicate we could reduce $N_{\text{IS}}$ by a
lot, down to $10$ in this specific setting for $n = 50$.

The comparison for varying $n$ yields interesting insight. The exact inflection
point may change depending on the problem, but the hardest problems for
saddlepoint Monte Carlo concerns cases where both uniform and normal sampling
do not work, which is the case where $X$ may have a complex or high-dimensional
structure, but $AX$ may not be well approximated by a saddlepoint method. In
this setting of ecological inference with uniform probabilities $p$, this is
when $n = 5$, but this would change as a function of $p$. As a rule of thumb
for practitioners, we consider it worthwhile to assess whether saddlepoint
Monte Carlo is useful in a given problem when $\min_i n p_{i.} \geq 3, \min_j
	np_{.j} \geq 3$. Study 3 aims to more precisely study how saddlepoint Monte
Carlo may behave with different choices of $p$.

\subsection{Efficiency given the characteristics of $AX$}

As described earlier, saddlepoint Monte Carlo works well (meaning it requires
few simulations to reach low variance) when saddlepoint approximations work
well. This is true when uniform rate approximation results hold, which concerns
near-log-concave distributions and near-Gaussian distributions, among others.
However, a key point is that the quality of the approximation depends on the
distribution of $AX$, and not of $X$. If $X$ is near-Gaussian, this will imply
that $AX$ is also near Gaussian, but it is not a necessary condition.

To demonstrate this fact in the context of ecological inference, let us
introduce two families of probabilities $p$ with $X \sim \mathcal{M}(n, p)$.
Representing $p$ in a $I \times I$ matrix, we denote $D_{m}$ as the $m-$th
diagonal $I \times I$ matrix ($m = 0$ corresponds to the standard diagonal). We
then define, with $\alpha$ a coefficient, the matrix
\begin{equation}
	p_\alpha^1 = \frac{1}{C}\sum_{m=-I/2}^{I/2} \alpha ^ m D_m,
\end{equation}
with $C$ a proportionality constant. We denote this family as ``type 1'', and $\alpha$ as an asymmetry coefficient. When $\alpha = 1$, this corresponds to the uniform probability matrix. When $\alpha > 1$ and the matrix is of size $3 \times 3$, the ratio between the biggest value of $p_\alpha^1$ and the smallest is $\alpha^2$. Increasing $\alpha$ thus leads to a more asymetric vector $X$ and distance from its Gaussian approximation. However, the marginal probabilities $A p_\alpha^1$ do not change as a function $\alpha$, which means that $AX$ will remain close to its Gaussian approximation.

As a point of comparison, we introduce a second probability family,
$(p^2_\alpha)_\alpha$, defined as, with $\tilde D_m$ an $I \times I$ matrix
with only its $m-$th row's coefficients equal to one (and the rest null),
\begin{equation}
	p_\alpha^2 = \frac{1}{C}\sum_{m = 0}^{I} \alpha^m \tilde D_m,
\end{equation}
with $C$ a proportionality constant. The idea is very similar here, with $\alpha = 1$ meaning $p_\alpha^2$ is the uniform probability matrix, and increasing $\alpha$ leading to more asymmetrical $AX$.

\begin{figure}
  \includegraphics{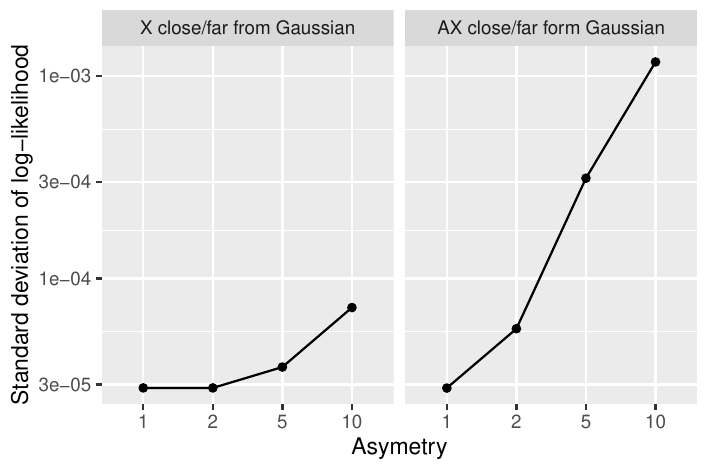}
  \caption{Comparison of the standard error of $\log \hat \delta ^{\text{Norm,
    tilt}}_{\theta, b}$ where $X \sim \mathcal{M}(n, p)$ and $n = 3000$, for
    choices of $p$ corresponding to $(p_\alpha^1)_{\alpha \geq 1}$ ($X$
    close/far from Gaussian, left pane) and $(p_\alpha^2)_{\alpha \geq 1}$
    ($AX$ close/far from Gaussian, right pane), with different asymmetry
    coefficient $\alpha = 1, 2, 5, 10$. Each experiment is done with $K = 200$
    observed units, $N_{\text{Sim}}=1000$ simulations, the standard error is
    computed for each unit over $200$ experiments and then averaged across
    units, and the $b_k$'s are simulated as $b = AX_k$ with $X_k \sim
  \mathcal{M}(n, p)$.}
  \label{fig:sparsity}
\end{figure}

\Cref{fig:sparsity} presents the standard deviation of the log-likelihood
computed for different values of asymmetry coefficients in each family of
probabilities $p^{1,2}_\alpha$, all else being held constant. $n$ is
voluntarily chosen high to showcase that drastic changes in efficiency happen
even in cases where one could think the Gaussian approximation of $X$ would
hold well. One observation can be made from this. Despite being constructed in
a rather similar manner, the standard error does not evolve the same way at all
for $(p^1_\alpha)$ and $(p^2_\alpha)$. In the first case, where $X$ is more or
less sparse but the marginal probabilities on $AX$ do not change and it remains
close to a Gaussian, the standard deviation of the log-likelihood is only
multiplied by $2.5$ when the asymetry coefficient $\alpha$ is changed from 1 to
10. In the second case, where marginal probabilities on $AX$ are changed, the
standard deviation explodes when $\alpha$ increases, with multiplication by a
factor of $41$ when $\alpha$ is changed from 1 to 10. This empirical study
illustrates why it is rather properties of $AX$ that matter when employing
saddlepoint approximation instead of just $X$. It is possible to have $X$ far
from a Gaussian and yet the standard deviation of the log-likelihood to remain
very small.

This simulation exercise has multiple consequences and interpretations for
ecological inference. First, it means that in settings where one studies a
large ecological table (and so exact MCMC schemes become too costly) with high
asymetries between coefficients (because of high dependence across categories),
saddlepoint Monte Carlo can work even if $X$ may not be approximated by a
Gaussian. Second, however, the condition for that is that all marginal
probabilities and counts remain not too low. For example, saddlepoint Monte
Carlo will work extremely well for studying voting behavior for candidates with
a decent vote share but where voice carryover is highly asymmetrical or
heterogeneous across constituencies, but not as well when studying candidates
very very small vote shares. In this case, it will be required to increase
$N_{\text{Sim}}$ to reduce the standard deviation of the log-likelihood,
leading to an increased computational budget.

\subsection{Tail behavior of saddlepoint Monte Carlo}

Finally, we aim to illustrate that the advantage of traditional saddlepoint
methods, namely that they exhibit a very low relative error when estimating
tail probabilities, transfers to saddlepoint Monte Carlo. This is important in
our context as we aim to use saddlepoint Monte Carlo to train statistical
models, which means that we might need to provide estimates for tail
probabilities either at the beginning of training, or when the model being
trained is misspecified. This property is crucial in the sense that it makes
training models with saddlepoint Monte Carlo much more stable than with other
importance sampling approaches for characteristic functions inversions.

\begin{figure}[h!]
	\includegraphics{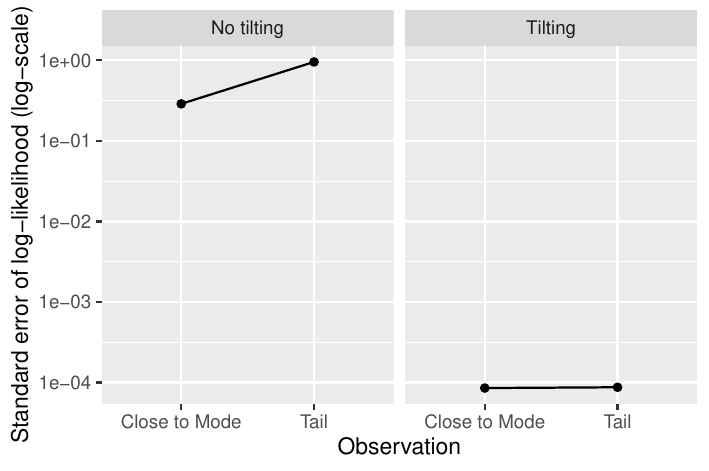}
	\caption{Comparison of the standard error of $\log \hat \delta ^{\text{Norm, tilt}}_{\theta, b}$ (left) and $\log \hat \delta ^{\text{Norm}}_{\theta, b}$ (right) with $X \sim \mathcal{M}(n, p)$, $n = 1000$, $p$ uniform, for observations $b$ that are either close to the mode (simulated with $p$ uniform) or far in the tail (simulated with $p$ = $p_\alpha^1$ with sparsity parameter $\alpha = 3$. $K = 200$ units are simulated once, the standard error is computed for each unit across $200$ experiments, each with $N_{\text{Sim}} = 1000$ simulations, before averaging across units.}
	\label{fig:tail_behavior}
\end{figure}

For the first time in \cref{fig:tail_behavior}, we will estimate
probabilities of $AX$ for vectors $b_k$ that are not simulated for the
distribution of $X$. We estimate the probability $\pi(AX=b)$ when assuming $X
	\sim \mathcal{M}(n, p^{\text{Unif}})$ with $p^{\text{Unif}}$ uniform, for two
choices of $b$. Vector $b$ is either simulated as $b = AX$ with $X \sim
	\mathcal{M}(n, p^{\text{Unif}})$, as before (setting ``close to the mode''), or
as $b = AX$ with $X \sim \mathcal{M}(n, p^1_\alpha), \alpha = 3$ (setting
``tail'', or far away from the mode). $n = 1000$ is voluntarily chosen high so
that the misspecification for the ``tail'' setting is very high. For each case,
we compare the tilted and the non-tilted Gaussian IS estimator.

Two results can be drawn from \cref{fig:tail_behavior}. First, we find
another confirmation that tilting drastically reduces the standard deviation of
the likelihood, here by a factor of at least $10 ^3$, which implies a
drastically smaller computational budget. Second, we observe that the standard
deviation for the non-tilted estimator is three times larger when estimating a
probability in the tail compared to close to the mode of the distribution. In a
practical context, this would mean that 9 times more simulations are needed
when estimating a probability in the tail, and since it is not really possible
to know if a certain point is in the tail before running the estimator, it
would require to increase the simulations by such a factor to avoid an unstable
training. On the other hand, for the tilted estimator, the standard error of
the log-likelihood is unchanged across observations close to the mode or in the
tail. The other choices for $p$ in our tests to the same conclusion, even if
the relative change for the non-tilted estimator may change from one
application to the other.

\section{Data preparation work}\label{app:data}

Some data preparation work is required so to work on voting stations (``bureaux de
vote'') for all of France at scale. First, we need to link voting stations across
the two election rounds. This requires correcting for the fact that some people
may move between the two rounds, leading to some small size differences; and
sometimes, for larger restructuring. Our decision is as follows: if a voting
station has a difference of more than 50 registered voters, it is discarded;
otherwise, electors are added to the ``Abstention'' category of the round for
which there is the smallest number of registered voters so that both rounds
match. In addition, some voting stations exhibit very non-standard behavior, such as
100\% of abstention voters, which we interpret as consequences of a judge ordering that votes be voided in the voting station; we discard such voting stations.

In addition, for ease of inference, we merge very small voting stations. Very
small voting stations of size less than 70 voters are merged at the level of the
department for the 2007 presidential election; and at the level of the
constituency for the 2022 and 2024 elections (this concerns around $0.1$\% of
voters in all cases). Finally, some voting stations may exhibit other non-standard
behavior which are spotted during inference (as they drive the far majority of
the variance), removed, and then qualitatively evaluated for confirmation. This
concerns 2 to 5 stations out of the 60\,000. We assume this is because they are
completely out of distribution: the situation could maybe be addressed by
adopting a more flexible model, like a hierarchical one (but which falls
outside of the scope of this paper).

In addition, as is common in the EI literature, we merged small candidates
together, to facilitate inference at moderate cost. Less merging would be
required with a more flexible model, or with smaller datasets, but we are
already able to explore tables that are considerably larger than the 2 by 2
case. Analysis for the 2024 legislative elections showcases even larger tables.
This is done manually for the presidential elections, and automatically with
the legislative elections, where all candidates obtaining less than 5\% of
votes are merged.

\subsection{Context variable pre-processing}

The density variable was pre-processed as follows: divide the number of
individuals by the surface area of the city (in squared kilometers), take the
logarithm (all values below 0 are then set to 0), and then normalize (center
and divide by standard deviation). One station change of the resulting variable
may be interpreted as one-standard deviation change of the density expressed in
log-scale.

\subsection{Precise field}

For 2007 presidential election, inference is performed on all voting stations
except minor restrictions listed below. For the 2022 presidential elections,
the inference is only performed on stations that could be merged with the Insee's
file ``Terrorities repertory'' given the identifiers, which represent 96\% of
voters. In addition, constituencies of French people not residing in France and
the Paris 1 constituency (which contains prisoners) are dropped out
because of how they stand out in terms of size. For both presidential
elections, voting stations in which O voters voted for any
of the main available options (Sarkozy, Royal, Bayrou, Le Pen; Sarkozy, Royal;
Le Pen, Macron, Melenchon; Le Pen, Macron) were removed (this concerns 9 voting
stations in 2007 and 18 in 2022). For the 2024 legislative elections, inference is
performed on all constituencies except for French people not residing in
France. When matched considering the aspect of the specific party of candidates
of the left, only constituencies in metropolitan France are considered.

\section{Two additional constituencies}

\begin{table}[h]
	\tiny
	\centering
	\begin{tabular}{lrrrrrrrrrrrr}
\toprule
 & \multicolumn{3}{r}{abstention} & \multicolumn{3}{r}{VERBRUGGHE RN} & \multicolumn{3}{r}{DARMANIN ENS} & \multicolumn{3}{r}{other} \\
 & .05 & .5 & .95 & .05 & .5 & .95 & .05 & .5 & .95 & .05 & .5 & .95 \\
\midrule
abstention & 0.84 & 0.86 & 0.88 & 0.02 & 0.03 & 0.04 & 0.07 & 0.09 & 0.11 & 0.01 & 0.02 & 0.03 \\
VERBRUGGHE RN & 0.04 & 0.07 & 0.09 & 0.86 & 0.89 & 0.91 & 0.01 & 0.02 & 0.05 & 0.01 & 0.02 & 0.04 \\
MORTREUX UG & 0.22 & 0.27 & 0.32 & 0.01 & 0.02 & 0.03 & 0.61 & 0.66 & 0.71 & 0.03 & 0.05 & 0.08 \\
DARMANIN ENS & 0.02 & 0.03 & 0.04 & 0.03 & 0.05 & 0.07 & 0.89 & 0.91 & 0.93 & 0.00 & 0.01 & 0.02 \\
other & 0.11 & 0.20 & 0.31 & 0.21 & 0.33 & 0.46 & 0.20 & 0.34 & 0.49 & 0.04 & 0.12 & 0.20 \\
\bottomrule
\end{tabular}

	\begin{tabular}{lrrrrrrrrrrrrrrr}
\toprule
 & \multicolumn{3}{r}{abstention} & \multicolumn{3}{r}{blanc nul} & \multicolumn{3}{r}{LEAUMENT UG} & \multicolumn{3}{r}{AMAND RN} & \multicolumn{3}{r}{other} \\
 & .05 & .5 & .95 & .05 & .5 & .95 & .05 & .5 & .95 & .05 & .5 & .95 & .05 & .5 & .95 \\
\midrule
abstention & 0.85 & 0.87 & 0.88 & 0.00 & 0.00 & 0.01 & 0.11 & 0.12 & 0.14 & 0.00 & 0.01 & 0.01 & 0.00 & 0.00 & 0.00 \\
LEAUMENT UG & 0.03 & 0.05 & 0.09 & 0.00 & 0.00 & 0.01 & 0.90 & 0.94 & 0.97 & 0.00 & 0.01 & 0.02 & 0.00 & 0.00 & 0.00 \\
MONET ENS & 0.05 & 0.09 & 0.14 & 0.34 & 0.38 & 0.41 & 0.18 & 0.23 & 0.27 & 0.26 & 0.31 & 0.35 & 0.00 & 0.00 & 0.00 \\
AMAND RN & 0.06 & 0.09 & 0.12 & 0.01 & 0.02 & 0.04 & 0.01 & 0.03 & 0.06 & 0.83 & 0.86 & 0.90 & 0.00 & 0.00 & 0.00 \\
other & 0.19 & 0.27 & 0.34 & 0.05 & 0.09 & 0.13 & 0.37 & 0.44 & 0.53 & 0.13 & 0.20 & 0.27 & 0.00 & 0.00 & 0.00 \\
\bottomrule
\end{tabular}

    \caption{Second round behavior in Calvados 6 and  Somme 1.
      Probabilities of voting in the second round (columns) conditional on
      first-round behavior (rows), with median posterior probability and 90\%
      credibility intervals. Political labels meaning: RN: far-right, UG: left,
    ENS \& DVC: centre.}
	\label{fig:other_constituencies}
\end{table}

\bibliographystyle{apalike}

\end{document}